\documentclass[amsmath,amssymb,12pt, eps,amsfonts]{article}
\usepackage{amsfonts,amssymb,amsmath,epsfig,epic,color}
\usepackage{bm}
\newcommand{\be}{\begin{equation}}
\newcommand{\ee}{\end{equation}}
\newcommand{\beqs}{\begin{eqnarray}}
\newcommand{\eeqs}{\end{eqnarray}}

\begin{document}
\title{ \bf Non-relativistic Lee Model \\  in Two Dimensional Riemannian Manifolds}

\author{\centerline {Fatih Erman$^1$, O. Teoman Turgut$^{1,\,2}$}
\\\and
 {\scriptsize{$^1$
Department of Physics, Bo\u{g}azi\c{c}i University, Bebek, 34342,
\.Istanbul, Turkey}}
\\\and
{\scriptsize{$^2$Feza G\"{u}rsey Institute, Kuleli Mahallesi,
\c{S}ekip Ayhan \"{O}z{\i}\c{s}{\i}k Caddesi, No: 44, Kandilli,
34684, \.Istanbul, Turkey}}
\\
{\scriptsize{E-mail: fatih.erman@gmail.com,
turgutte@boun.edu.tr}}}

\date{\scriptsize{\textsc{\today}}}

\maketitle


\begin{abstract}

This work is a continuation of our previous work (JMP,
\textbf{48}, 12, pp. 122103-1-122103-20, 2007), where we
constructed the non-relativistic Lee model in three dimensional
Riemannian manifolds. Here we renormalize the two dimensional
version by using the same methods and the results are shortly
given since the calculations are basically the same as in the
three dimensional model. We also show that the ground state energy
is bounded from below due to the upper bound of the heat kernel
for compact and Cartan-Hadamard manifolds. In contrast to the
construction of the model and the proof of the lower bound of the
ground state energy, the mean field approximation to the two
dimensional model is not similar to the one in three dimensions
and it requires a deeper analysis, which is the main result of
this paper.

\end{abstract}


\section{Introduction}

The Lee model was originally introduced in \cite{lee} as an
exactly soluble (in principle eigenstates and eigenvalues can be
exactly found) and a renormalizable model that describes the
interaction between a relativistic neutral bosonic field ``pions"
and two neutral fermionic fields ``nucleons". It is assumed that
the nucleon can exist in two different intrinsic states. The
particle corresponding to the Bose field is called $\theta$ and
the particles corresponding to the intrinsic states of the nucleon
are called $V$ and $N$ particles. The fermionic field
corresponding to the $V$ and $N$ particles are assumed to be
spinless for simplicity. Only allowable process is given by
\be V \leftrightharpoons N + \theta \ee
and the following process is not allowed
\be N \leftrightharpoons V + \theta \,,\ee
which makes the model rather simple. Although this model is not
realistic, the important features of nucleon-pion system can be
understood in a relatively simple way and one can get rid of the
infinities without applying perturbation theory techniques.
Moreover, the complete non-relativistic version of this model that
describes one heavy particle sitting at some fixed point
interacting  with a field of non-relativistic bosons is as
important as its relativistic counterpart. It is much simpler than
its relativistic version because only an additive renormalization
of the mass difference of the fermions is necessary. It has been
studied in a textbook by Henley and Thirring for small number of
bosons from the point of view of scattering matrix \cite{Thirring}
and there are various other approaches to the model
\cite{trubatch,nickle,dittrich,fuda,bender,morris,marshall}. It is
possible to look at the same problem from the point of view of the
resolvent of the Hamiltonian in a Fock space formalism with
arbitrary number of bosons (in fact there is a conserved quantity
which allows us to restrict the problem to the direct sum of $n$
and $n+1$ boson sectors). This is achieved in a very interesting
unpublished paper by S. G. Rajeev \cite{rajeevbound},
 in which a new non-perturbative formulation of renormalization for some models
 with contact interactions has been
proposed.

This paper is a natural continuation of our previous work
\cite{ermanturgut2} and we extend the three dimensional model
constructed there to the two dimensional one. In
\cite{ermanturgut2}, we discussed the non-relativistic Lee model
on three dimensional Riemannian manifolds by following
\cite{rajeevbound} and renormalized the model by the help of heat
kernel techniques with the hope that one may understand the nature
of renormalization on general curved spaces better. In fact, the
idea developed in \cite{rajeevbound} has also been applied to
point interactions in \cite{altunkaynak etal,ermanturgut,caglar}
and to the relativistic Lee model in \cite{kaynakturgut}. In this
paper, we are not going to review the ideas developed in
\cite{rajeevbound} and \cite{ermanturgut2}. Instead, we recommend
the reader to read through these papers. The construction of the
model is exactly the same as the one in three dimensions. It is
again based on finding the resolvent of the regularized
Hamiltonian $H_\epsilon$ and showing that a well-definite finite
limit of the resolvent exists as $\epsilon \rightarrow 0^+$
(called renormalization) with the help of heat kernel. We then
prove that the ground state energy for a fixed number of bosons is
bounded from below, using the upper bound estimates of heat kernel
for some classes of Riemannian manifolds. Finally, we study the
model in the mean field approximation for compact and non-compact
manifolds separately. Although the construction of the model and
lower bound of the ground state energy are based on the same
calculations as in three dimensions, \textit{the mean field
approximation in two dimensions requires a deeper analysis} as we
shall see, which is the main result of the present work.

The paper is organized as follows. In the first part, we give a
short construction of the model and show that the renormalization
can be accomplished on two dimensional Riemannian manifolds. Then,
we prove that there exists a lower bound on the ground state
energy. Finally, the model is examined in the mean field
approximation. In appendix,  we prove an inequality which we use
in the mean field approximation.

\section{Construction of the Model in Two Dimensions}

In this section, we give a brief summary of the construction of
the model in two dimensions since it has been basically done in
\cite{ermanturgut2} for the three dimensional model. We start with
the regularized Hamiltonian of the non-relativistic Lee model on
two dimensional Riemannian manifold $(\mathcal{M},g)$ with a
cut-off $\epsilon$. In natural units ($\hbar=c=1$), the
regularized Hamiltonian on the local coordinates $x=(x_1,x_2) \in
\mathcal{M}$ is
\begin{equation}
H^{\epsilon} =H_0 + H_{I,\epsilon} \;,
\end{equation}
where
\begin{equation}
H_0=\int_{\mathcal{M}} \mathrm{d}_{g}^{2} x \;
\phi^\dag_{g}(x)\left(-{1\over 2m}\nabla^2_g+m \right)\phi_g(x)
\;,
\end{equation}
\begin{equation}
H_{I,\epsilon}=\mu (\epsilon) \left( {1-\sigma_3\over 2} \right)+
\lambda \int_{\mathcal{M}} \mathrm{d}_{g}^{2} x \;
K_\epsilon(x,a;g)\left(\phi_g(x)\, \sigma_-+\phi^\dag_{g}(x) \,
\sigma_+ \right) \;. \end{equation}
Here $\phi^{\dag}_{g}(x)$, $\phi_{g}(x)$ is the bosonic
creation-annihilation operators defined on the manifold with the
metric structure $g$ and $\lambda$ is the coupling constant, and
$x,y$ refers to points on the manifold $\mathcal{M}$. Also, $
K_\epsilon(x,a;g)$ is the heat kernel on a Riemannian manifold
with metric structure $g$ and it converges to the Dirac delta
function $\delta_g (x,a)$ around the point $a$ on $\mathcal{M}$ as
we take the limit $\epsilon \to 0^+$. For simplicity, we have
changed the notation for the heat kernel to $K_s(x,y;g)$ instead
of writing $K_{s/2m}(x,y;g)$ which was used in
\cite{ermanturgut2}.  We also assume stochastic completeness, that
is
\begin{equation}
\int_{\mathcal M} \mathrm{d}_{g}^{2} x \; K_s(x,y;g)=1 \;.
\end{equation}
Similar to the flat case,
$\mu(\epsilon)$ is defined as a bare mass difference between the
$V$ particle (neutron) and the $N$ particle (proton). Although the
number of bosons is not conserved in the model, one can derive
from the equations of motion that there exists a conserved
quantity
\begin{equation} \nonumber
Q = - \left({1+ \sigma_3 \over 2}\right) + \int_{\mathcal{M}}
\mathrm{d}_{g}^{2} x \; \phi^{\dag}_g(x) \phi_g(x) \;.
\end{equation}
Therefore, we can express the regularized Hamiltonian as a
$2\times 2$ block split according to $\mathbb{C}^2$:
\begin{equation} H^{\epsilon} -E =
\left(%
\begin{array}{cc}
  H_0-E & \lambda \int_{\mathcal{M}} \mathrm{d}_{g}^{2} x \; K_\epsilon(x,a;g)\, \phi^\dag_g(x) \\
   \lambda \int_{\mathcal{M}} \mathrm{d}_{g}^{2} x \; K_\epsilon(x,a;g)\, \phi_g(x) & H_0- E
   +\mu(\epsilon)
\end{array}%
\right)\;.
   \end{equation}
Then, the regularized resolvent of this Hamiltonian in two
dimensions is
\begin{equation}
R^{\epsilon}(E)= {1 \over H^{\epsilon} -E}= \left(%
\begin{array}{cc}
  \alpha_\epsilon & \beta^{\dagger}_{\epsilon} \\
  \beta_{\epsilon} & \delta_{\epsilon} \\
\end{array}%
\right) =  \left( \begin{array}{cc}
  a_\epsilon & b^\dagger_\epsilon \\
 b_\epsilon & d_\epsilon \\
\end{array}\right)^{-1} \;,
\end{equation}
where
\begin{eqnarray}
\alpha_\epsilon&=&{1\over H_0-E}+{1\over H_0-E} \;
b^{\dag}_{\epsilon} \; \Phi_\epsilon^{-1} (E) \; b_\epsilon \;
{1\over H_0-E}\cr \beta_\epsilon &=&- \Phi^{-1}_\epsilon(E)\;
b_\epsilon \; {1\over H_0-E}\cr \delta_\epsilon
&=&\Phi^{-1}_\epsilon(E)\cr b_\epsilon &=&\lambda
\int_{\mathcal{M}} \mathrm{d}_{g}^{2} x \;
K_\epsilon(x,a;g)\,\phi_g(x) \;.
\end{eqnarray}
Here $E$ should be considered as a complex variable. Most
importantly, the operator $\Phi_\epsilon(E)$, called principal
operator, is given as
\begin{equation}
\Phi_\epsilon(E)=H_0-E+\mu(\epsilon)-\lambda^2
\int_{\mathcal{M}^2} \mathrm{d}_{g}^{2} x \, \mathrm{d}_{g}^{2} y
\; K_\epsilon(x,a;g) K_\epsilon(y,a;g)\; \phi_g(x){1\over
H_0-E}\phi^\dagger_g(y) \;.\label{phi_e}
\end{equation}
After performing normal ordering of this operator, we get
\begin{eqnarray}
& & \Phi_\epsilon(E)= H_0-E-\lambda^2 \int_{\epsilon/2}^\infty
\mathrm{d} s \int_{\mathcal{M}^2} \mathrm{d}_{g}^{2} x \,
\mathrm{d}_{g}^{2} y \; K_{s}(x,a;g)K_{s}(y,a;g) \cr & &
\hskip-1.2cm \times \; \phi^{\dag}_{g}(x)
e^{-(s-\epsilon/2)(H_0+2m-E)}\phi_{g}(y) +
 \mu+\lambda^2  \int_\epsilon^\infty \mathrm{d} s \;
K_{s}(a,a;g)\left[e^{-s(m-\mu)}-e^{-(s-\epsilon)(H_0+m-E)}\right]\;.
\end{eqnarray}
If we choose $\mu(\epsilon)$ as
\begin{eqnarray} \label{mueps}
\mu(\epsilon) = \mu + \lambda^2 \int_{\epsilon}^{\infty}
\mathrm{d} s \; K_{s} (a,a;g)\, e^{-s(m-\mu)}\;,
\end{eqnarray}
where $\mu$ is defined as the physical energy of the composite
state which consists of a boson and the attractive heavy neutron
at the center $a$, and take the limit $\epsilon\to 0^+$, we obtain
\begin{eqnarray}
\Phi(E)&=&H_0-E+\mu+\lambda^2 \int_0^\infty \mathrm{d} s \;
K_{s}(a,a;g)\left[e^{-s(m-\mu)} -e^{-s(H_0+m-E)}\right]\cr & \ &
\hskip-2cm -\lambda^2 \int_0^\infty \mathrm{d} s
\int_{\mathcal{M}^2} \mathrm{d}_{g}^{2} x \, \mathrm{d}_{g}^{2} y
\; K_{s}(x,a;g)K_{s}(y,a;g) \, \phi^\dag_g(x)
e^{-s(H_0+2m-E)}\phi_g(y) \;. \label{renormalized principal}
\end{eqnarray}
This is the renormalized form of the principal operator so that we
have a well-defined explicit formula for the resolvent of the
Hamiltonian in terms of the inverse of the principal operator
$\Phi^{-1}(E)$. As we will elaborate in more detail later on, the
bound states arise from the roots of the equation
\be \Phi(E) |\Psi \rangle =0 \;, \ee
corresponding to the poles in the resolvent.
Hence, the principal operator determines the bound state spectrum.

\section{A Lower Bound on the Ground State Energy}
\label{A Lower Bound on the Ground State Energy for Two and
Three Dimensions}

Following exactly the same method developed for three dimensions
\cite{ermanturgut2}, we again find the upper bound of the norm of
the operator $\tilde{U}'(E)$ as
\begin{eqnarray} & &
||\tilde{U}'(E)|| \leq  n \; { \lambda^2 \, \Gamma(2)\over
\Gamma(1/2)^2} \int_0^\infty \mathrm{d} s \; s \; e^{-s(nm+\mu-E)}
\int_0^1 \mathrm{d}u_1 \int_0^1 \mathrm{d}u_2 \int_0^1
\mathrm{d}u_3 \, { \delta(u_1+u_2+u_3-1) \over (u_1 \, u_2)^{1/2}}
 \cr
& & \hspace{4cm} \times  \left[ K_{2s(u_1+u_3)}(a,a;g)
\right]^{1/2} \left[ K_{2s(u_2+u_3)}(a,a;g) \right]^{1/2}\;,
\label{normU} \end{eqnarray}
where $\Gamma$ denotes the gamma function. For each class of
manifolds, there are different upper bounds on the heat kernel so
we will consider them separately. We will first consider
Cartan-Hadamard manifolds. The diagonal upper bound of the heat
kernel for two dimensional Cartan-Hadamard manifolds is given as
\cite{grigoryan,mckean},
\begin{equation} \label{cartanestim}
K_{s} (x,x;g) \leq {C \over (s/2m)} \;,
\end{equation}
for all $x\in \mathcal{M}$, $s>0$, and $C$ is a constant. Using
(\ref{cartanestim}) and performing $u_3$ integral in
(\ref{normU}), we get
\beqs & & ||\tilde{U}'(E)||  \leq  n \; C m \; { \lambda^2 \,
\Gamma(2)\over \Gamma(1/2)^2} \int_0^\infty \mathrm{d} s \;
e^{-s(nm+\mu-E)} \cr & & \hspace{4cm}\times \int_0^1 \mathrm{d}u_1
{1 \over u_{1}^{1/2}} {1 \over (1-u_1)^{1/2}} \int_{0}^{1-u_1}
\mathrm{d}u_2 \,  {1 \over u_{2}^{1/2}} {1 \over (1-u_2)^{1/2}}
\cr & & \hspace{1.8cm} \leq  n \; C m \; { \lambda^2 \,
\Gamma(2)\over \Gamma(1/2)^2} \int_0^\infty \mathrm{d} s \;
e^{-s(nm+\mu-E)} \Bigg[\int_0^1 \mathrm{d}u {1 \over u^{1/2}} {1
\over (1-u)^{1/2}} \Bigg]^2 \;. \eeqs
Evaluating the integrals give the following result
\beqs ||\tilde{U}'(E)|| \leq n \; C m \pi \;  \lambda^2 \,
\Gamma(2) (nm + \mu -E)^{-1} \;. \eeqs
Then, the inequality above implies a lower bound for the ground
state energy
\begin{equation}
E_{gr} \geq n m  + \mu  - n \tilde{C} \lambda^2 m
\label{energycartan}\;,
\end{equation}
where %
\begin{equation}
\tilde{C} = C \pi \Gamma(2) \;.
\end{equation}
The diagonal upper bound of the heat kernel for two dimensional
compact manifolds with Ricci curvature bounded from below by $-K
\geq 0$ is given by \cite{liyau,wang},
\begin{equation}
K_s(x,x;g) \leq {1 \over V(\mathcal{M})} + A(s/2m)^{-1} \;,
\label{compact upper bound heat kernel}
\end{equation}
where $A$ depends on the diameter of the manifold,
$d(\mathcal{M})$, the lower bound of the Ricci curvature $K$, and
the volume of the manifold $V(\mathcal{M})$. Then we similarly
obtain
\begin{eqnarray} & &
||\tilde{U}'(E)|| \leq  n \;  {\lambda^2 \, \Gamma(2)\over
\Gamma(1/2)^2} \int_0^\infty \mathrm{d} s \; s \; e^{-s(nm+\mu-E)}
\Bigg\{ \int_0^1 {\mathrm{d}u_1 \over (u_1)^{1/2}} \Bigg[ {1 \over
V(\mathcal{M})^{1/2}} \cr & & + A^{1/2} (s(1-u_1)/m)^{-1/2}
 \Bigg] \int_0^{1-u_1}
{\mathrm{d}u_2 \over (u_2)^{1/2}}
   \left[ {1 \over V(\mathcal{M})^{1/2}} + A^{1/2} (s(1-u_2)/m)^{-1/2}
\right] \Bigg\} \;.
\end{eqnarray}
One can even simplify the integrals, that is, the upper bound of
the $u_2$ integral is replaced with 1 and the square roots of the
sums are replaced with the sums of the square roots at the cost of
getting less sharp bound on the norm of $\tilde{U}'(E)$.
Integrating with respect to $u_1$, $u_2$ and $s$, we have
\begin{eqnarray}
||\tilde{U}'(E)|| \leq n \; {\lambda^2 \, \Gamma(2)\over
\Gamma(1/2)^2} \bigg[{4 \over V(\mathcal{M})}{1 \over (n m + \mu
-E)^2} + {4 A^{1/2} m^{1/2} \pi^{1/2} \Gamma({3\over 2})
\Gamma({1\over 2}) \over V(\mathcal{M})^{1/2} (n m + \mu -E)^{{3
\over 2}}} + {A m \pi \Gamma({1 \over 2})^2 \over (n m + \mu -E)}
 \bigg] \cr \;.
\end{eqnarray}
In order to get an explicit solution of this inequality, let us
put a further natural assumption $nm + \mu - E > \mu$. Then, we
find
\begin{eqnarray}
||\tilde{U}'(E)|| \leq n \; {\lambda^2 \, \Gamma(2)\over
\Gamma(1/2)^2} \bigg[{4 \over V(\mathcal{M}) \mu} + {4 A^{1/2}
m^{1/2} \pi^{1/2} \Gamma({3\over 2}) \Gamma({1\over 2}) \over
\mu^{1 \over 2} V(\mathcal{M})^{1/2}} + A m \pi \Gamma(1/2)^2
 \bigg] {1 \over (n m + \mu - E)} \;.
\end{eqnarray}
Now if we impose the strict positivity of the principal operator,
we obtain
\begin{equation}
E_{gr} \geq n m + \mu - n \lambda^2 F \label{energyh4}\;,
\end{equation}
where
\beqs & & F = {\Gamma(2) \over \Gamma(1/2)^2} \bigg[{4 \over
V(\mathcal{M}) \mu} + {4 A^{1/2} m^{1/2} \pi^{1/2} \Gamma({3\over
2}) \Gamma({1\over 2}) \over \mu^{1 \over 2} V(\mathcal{M})^{1/2}}
+ A m \pi \Gamma(1/2)^2
 \bigg] \;. \eeqs
Therefore, the lower bounds on the ground state energies for
different classes of manifolds (\ref{energycartan}) and
(\ref{energyh4}) are of almost the same form up to a constant
factor so the form of the lower bound has a general character.
From the general form of the lower bounds, we conclude that for
each sector with a fixed number of bosons, there exists a ground
state. However, in two dimensions, the ground state energy bound
that we have found diverges linearly as the number of bosons
increases whereas in the three dimensional case it diverges
quadratically. Therefore, unlike the three dimensional problem,
these estimates with our present analysis are good enough to prove
the existence of the thermodynamic limit in two dimensions. We
will now study large $n$ or the thermodynamic limit of the model
by a kind of mean field approximation, yet this requires a more
delicate analysis than the three dimensional case.

\section{Mean Field Approximation}
\label{Mean Field Approximation of the Model in Two Dimensions}

Before applying the mean field approximation, we first clarify one
point about this approximation, which has not been mentioned in
our previous work. It is well known that the residue of the
resolvent at its isolated pole $\mu$ is the projection operator
$\mathbb{P}_\mu$ to the corresponding eigenspace of the
Hamiltonian
\begin{equation} \label{projection resolvent}
\mathbb{P}_{\mu}= -{1 \over 2 \pi i} \oint_{\Gamma_{\mu}}
\mathrm{d} E \; R(E)\;,
\end{equation}
where $\Gamma_{\mu}$ is a small contour enclosing the isolated
eigenvalue $\mu$ in the complex energy plane \cite{simon}. For the
moment, let us consider only the first diagonal element of the
resolvent, $\alpha(E)$ and choose the contour enclosing the ground state energy
$E_{gr}$ which is a well-defined point on the real axis thanks to the bound given in the previous
section. Then, the above integral of $\alpha(E)$ gives the
projection to eigenspace $|\Psi_0 \rangle \langle \Psi_0|$
corresponding to the minimum eigenvalue.

From (\ref{renormalized principal}), it is easily seen that the
principal operator formally satisfies $\Phi^{\dag}(E)=
\Phi(E^{*})$. As will be shown elsewhere \cite{caglarermanturgut},
$\Phi(E)$ defines  a self-adjoint holomorphic family of type A
\cite{kato}, so that we can apply the spectral theorem for the
principal operator or inverse of it. Moreover, holomorphic
functional calculus applies to this family. Most importantly, it
can be shown that this family has  a common dense domain which
corresponds to (an operator closure of)  the domain of $H_0$ for
all sectors.
\beqs \Phi^{-1}(E) &=& \sum_{k} {1 \over \omega_k(E)}
\mathbb{P}_k(E) + \int_{\sigma} \mathrm{d} \omega(E)  \; {1 \over
\omega(E)} \mathbb{P}_{\omega}(E) \cr & =& \sum_{k} {1 \over
\omega_k(E)} |\omega_k(E) \rangle \langle \omega_k(E) | +
\int_{\sigma(\Phi)} \mathrm{d} \omega(E)  \; {1 \over \omega(E)}
|\omega(E) \rangle \langle \omega(E) | \;, \eeqs
where $\omega(E)$ ($\omega_k$ in the discrete case) and
$|\omega(E) \rangle$ ($|\omega_k(E) \rangle$ in the discrete case)
are the eigenvalues and the eigenvectors of the principal
operator, respectively. We assume that the principal operator has
discrete, assumed non-degenerate,  as well as continuous eigenvalues and the
bottom of the spectrum corresponds to an eigenvalue. The integrals
here are taken over the continuous spectrum $\sigma(\Phi)$ of the
principal operator (for simplicity, we write it formally, it
should be written more precisely as a Riemann-Stieltjies integral).
Due to Feynman-Hellman theorem, we have
\beqs & & {\partial \omega_k \over \partial E} = \langle
\omega_k(E) | {\partial \Phi(E) \over \partial E} | \omega_k(E)
\rangle \cr & & = - \Bigg( 1 + \lambda^2 \int_{0}^{\infty}
\mathrm{d} s \; s \; K_s(a,a;g) \left|\left| e^{-{s \over
2}(H_0-\mu + E)}| \omega_k(E) \rangle \right| \right|^2 \cr & & +
\lambda^2 \int_{0}^{\infty} \mathrm{d} s \; s \;  \left| \left|
e^{-{s \over 2} (H_0 - \mu + E)}\int_{\mathcal{M}}
\mathrm{d}_{g}^{2} x \; K_s(x,a;g) \phi_g(x) | \omega_k(E) \rangle
\right|\right|^2 \Bigg) < 0 \;, \label{derivative of eigenvalue
phi}\eeqs
by using  the positivity of the heat kernel. Note that the
operator valued distributions $\phi_g(x)$ becomes well defined by
taking a convolution with the heat kernel in the last term. The
bound state spectrum corresponds to the solutions of the zero
eigenvalues of the principal operator. $\omega_k(E)$'s flow with
$E$ due to (\ref{derivative of eigenvalue phi}), so that the
ground state corresponds to the zero of the minimum eigenvalue
$\omega_0(E)$ of $\Phi(E)$. Let us expand the minimum eigenvalue
$\omega_0(E)$ near the bound state energy $E_{gr}$
\begin{equation}
\omega_0(E) = \omega_0(E_{gr}) + (E- E_{gr}) {\partial \omega_0(E)
\over
\partial E}\bigg|_{E_{gr}}+ \cdots  = (E- E_{gr}) {\partial \omega_0(E)
\over
\partial E}\bigg|_{E_{gr}}+ \cdots  \;.
\end{equation}
Using this result and the residue theorem in (\ref{projection
resolvent}), we obtain
\beqs (H_{0}-E_{gr})^{-1} \phi^{\dag}(a) \left(- {\partial
\omega_0(E)\over
\partial E}|_{E_{gr}}  \right)^{-1}
|\omega_0(E_{gr}) \rangle \langle \omega_0(E_{gr}) | \phi(a)
(H_{0}-E_{gr})^{-1} \label{residue resolvent} \eeqs
There is no other  pole coming from $(H_0 -
E)^{-1}$ near $E_{gr}$ since we assume $E_{gr}<nm$, and no other terms for $k \neq 0$
contribute to the integral around $E_{gr}$. Let us assume that the
ground state eigenvector of the principal operator is
\be | \omega_0(E_{gr}) \rangle = \int_{\mathcal{M}^{n-1}}
\mathrm{d}_{g}^{2}x_1 \cdots \mathrm{d}_{g}^{2}x_{n-1} \;
\psi_{0}(x_1,\cdots, x_{n-1}) |x_1 \cdots x_{n-1} \rangle \;. \ee
By using the eigenfunction expansion of the creation and the
annihilation operators and their commutation relations, we shall
shift all creation operators $\phi^{\dag}_{g}(x)$ to the leftmost
\beqs & & \hskip-10cm {1\over H_0-E}
\phi^{\dag}_{g}(a)\phi^{\dag}_{g}(x_1)\cdots\phi^{\dag}_{g}(x_{n-1})
= \int_{\mathcal{M}^{n}} \mathrm{d}_{g}^{2} y_1 \cdots
\mathrm{d}_{g}^{2} y_n \; \phi^{\dag}_{g}(y_1)\cdots
\phi^{\dag}_{g}(y_n) \nonumber \\
\cr & \times \int_0^\infty \mathrm{d} s \; e^{-s(H_0-E + nm)} \,
K_{s}(y_1,a;g) K_{s}(y_2,x_1;g)\cdots K_{s}(y_{n},x_{n-1};g) \;,
\label{normalorderingcreation} \eeqs
and all annihilation operators $\phi_{g}(x)$ to the rightmost
\beqs & & \hskip-10cm
\phi_{g}(a)\phi_{g}(x_1)\cdots\phi_{g}(x_{n-1}) {1\over H_0-E} =
\int_{\mathcal{M}^{n}} \mathrm{d}_{g}^{2} y_1 \cdots
\mathrm{d}_{g}^{2} y_n \; \int_0^\infty \mathrm{d} s \; e^{-s(H_0-E + n m)}  \nonumber \\
\cr & \times \, K_{s}(y_1,a;g) K_{s}(y_2,x_1;g)\cdots
K_{s}(y_{n},x_{n-1};g) \phi_{g}(y_1)\cdots \phi_{g}(y_n) \;,
\label{normalorderingannihilation} \eeqs
which are the generalized versions of the equations we first used
in \cite{ermanturgut2}. Therefore, from the equation (\ref{residue
resolvent}), we read the state vector $|\Psi_0 \rangle$
\beqs & & |\Psi_0 \rangle = \int_{\mathcal{M}^{n}}
\mathrm{d}_{g}^{2} y_1 \cdots \mathrm{d}_{g}^{2} y_n \;
\Psi_{0}(y_1, \ldots, y_n) |y_1\cdots y_n \rangle \cr & & =
\int_{\mathcal{M}^{n}} \mathrm{d}_{g}^{2} y_1 \cdots
\mathrm{d}_{g}^{2} y_n  \; \int_{\mathcal{M}^{n-1}}
\mathrm{d}_{g}^{2} x_1 \cdots \mathrm{d}_{g}^{2} x_{n-1} \; {1
\over n} \sum_{\sigma \in (1 \cdots n)} \int_0^\infty \mathrm{d} s
\; e^{-s(nm-E_{gr})}
K_{s}(y_{\sigma(1)},a;g)\nonumber \\
\cr & & \hskip-1.2cm \times  K_{s}(y_{\sigma(2)},x_1;g)\cdots
K_{s}(y_{\sigma(n)},x_{n-1};g)\; \psi_{0}(x_1,\cdots,x_{n-1})
\left(- {\partial \omega_0(E)\over
\partial E}|_{E_{gr}}  \right)^{-1/2}
|y_{\sigma(1)}\cdots y_{\sigma(n)} \rangle \;, \label{psi0} \eeqs
where the sum runs over all cyclic permutations $\sigma$ of
$(123\ldots n)$. We will now make a mean field approximation to
this model. In standard quantum field theory, one expects that all
the bosons have the same wave function $u(x)$ for the limit of
large number of bosons $n\rightarrow \infty$ and the wave function
of the system has the product form of the one particle wave
functions. However, due to the singular structure of our problem,
the wave function in (\ref{psi0}) can not have a product form in
the large $n$ limit. In order to see this, we note that
$nm-E_{gr}$ is the crucial factor. If $nm-E_{gr}=O(n^{\alpha})$
where $\alpha$ is a positive exponent, we could get a
simplification. To demonstrate this, let  us  define
$nm-E_{gr}=(1-y[v])\chi -2m y[v]$, where $y[v]<1$. As we will see,
$\chi$ is what we typically estimate, and the variable $y[v]$ is
related to the scaled kinetic energy functional. Indeed, for
compact manifolds we will see that $\chi\approx n^{1/2}$ and
$y[v]\approx 0$. We scale $s=s'/[(1-y[v])\chi -2my[v]]$ and as
$n\rightarrow \infty$, all integrals of the heat kernels are
peaked around $y_{\sigma(k)}$. (This is clear from the property of
the heat kernel that $K_s(x,y;g) \rightarrow \delta_g(x,y)$ in the
sense of distributions as $s\rightarrow 0^+$ and also from the
stochastic completeness assumption). Then, all integrals of
$x_{\sigma(l)}$ are
\be \int_{\mathcal{M}}\mathrm{d}_{g}^{2} x_l \;
K_{s/[(1-y)\chi-2my]}(x_l ,y_{\sigma(l+1)})
\psi_0(x_1,\ldots,x_l,\ldots, x_{n-1}) \approx
\psi_0(x_1,\ldots,y_{\sigma(l+1)},\ldots, x_{n-1}) \;, \ee
for $l=1,\ldots,n-1$ as $n\rightarrow \infty$ and state $|\Psi_0
\rangle$ becomes
\beqs & & |\Psi_0 \rangle \approx \int_{\mathcal{M}^{n}}
\mathrm{d}_{g}^{2} y_1 \cdots \mathrm{d}_{g}^{2} y_n \; {1 \over
n} \sum_{\sigma \in (1 \cdots n)} \int_0^\infty \mathrm{d} s \;
e^{-s(nm-E_{gr})} K_{s}(y_{\sigma(1)},a;g)\nonumber \\
& & \psi_{0}(y_{\sigma(2)},\cdots,y_{\sigma(n)}) \left(- {\partial
\omega_0(E)\over
\partial E}|_{E_{gr}}  \right)^{-1/2}
|y_{\sigma(1)}\cdots y_{\sigma(n)} \rangle \;. \label{psi1} \eeqs
It is important to note that $|\Psi_0 \rangle$ is not in the
domain of $H_0$. In order to see this, it is sufficient to
consider the following term when we calculate $\langle \Psi_0 |
H_0 | \Psi_0 \rangle$
\beqs & & \int_{\mathcal{M}} \mathrm{d}^{2}_{g} x \;
\int_{0}^{\infty} \mathrm{d} s_1 \; e^{-s_1(nm-E_{gr})}
K_{s_1}(x,a;g) \Bigg[ \int_{0}^{\infty} \mathrm{d} s_2 \;
e^{-s_2(nm-E_{gr})} \left(-{1\over 2m} \right) \nabla_{g}^{2}
K_{s_2}(x,a;g) \Bigg] \cr & & \hskip-0.5cm = \int_{\mathcal{M}}
\mathrm{d}^{2}_{g} x \; \int_{0}^{\infty} \mathrm{d} s_1 \;
e^{-s_1(nm-E_{gr})} K_{s_1}(x,a;g) \Bigg[ \int_{0}^{\infty}
\mathrm{d} s_2 \; e^{-s_2(nm-E_{gr})} \bigg( -{\partial
K_{s_2}(x,a;g) \over
\partial s_2} \bigg) \Bigg] \;, \eeqs
where we have used the property that the heat kernel satisfies the
heat equation $-{1 \over 2m} \nabla_{g}^{2} K_{s}(x,a;g) +
{\partial K_s(x,a;g) \over \partial s} =0 $. After applying the
integration by parts to the $s_2$ integral and using the initial
condition for the heat kernel $K_s(x,a;g) \rightarrow
\delta_g(x,a)$ as $s\rightarrow 0^{+}$, we find
\beqs & & \int_{\mathcal{M}} \mathrm{d}^{2}_{g} x \;
\int_{0}^{\infty} \mathrm{d} s_1 \; e^{-s_1(nm-E_{gr})}
K_{s_1}(x,a;g) \Bigg[ \delta_g(x,a) \cr & & \hspace{6cm} -
(nm-E_{gr}) \int_{0}^{\infty} \mathrm{d} s_2 \;
e^{-s_2(nm-E_{gr})} K_{s_2}(x,a;g) \Bigg] \cr & & =
\int_{0}^{\infty} \mathrm{d} s_1 \; e^{-s_1(nm-E_{gr})}
K_{s_1}(a,a;g) - (nm-E_{gr}) \int_{0}^{\infty} \mathrm{d} s_1 \;
e^{-s_1(nm-E_{gr})} \cr & & \hspace{7cm} \times  \int_{0}^{\infty}
\mathrm{d} s_2 \; e^{-s_2(nm-E_{gr})} K_{s_1+s_2}(a,a;g) \eeqs
where we have used the semi-group property of the heat kernel
$\int_{\mathcal{M}} \mathrm{d}^{2}_{g} y \; K_{s_1}(x,y;g)
K_{s_2}(y,z;g) = K_{s_1+s_2}(x,z;g)$. After the change of
variables $u=s_1+s_2$ and $v=s_1-s_2$, we get
\beqs & & \int_{0}^{\infty} \mathrm{d} s_1 \; e^{-s_1(nm-E_{gr})}
K_{s_1}(a,a;g) -  (nm-E_{gr})\int_{0}^{\infty} \mathrm{d} u \; u
\; e^{-u(nm - E_{gr})} K_{u}(a,a;g) \;. \eeqs
The first term is divergent due to the short time asymptotic
expansion of the diagonal heat kernel for any Riemannian manifold
without boundary \cite{gilkey}
\be K_s(a,a;g) \sim {1 \over (4 \pi s /2m)} \sum_{k=0}^{\infty}
u_k(a,a)(s/2m)^{k} \;, \label{asymheat} \ee
where $u_k(a,a,)$ are scalar polynomials in curvature tensor of
the manifold and its covariant derivatives at point $a$. Similar
to the problem with point interactions on manifolds which we
studied in \cite{ermanturgut}, our problem here can also be
considered as a kind of self-adjoint extension since the wave
function $\Psi_0$ does not belong to the domain of the free
Hamiltonian. The self-adjoint extension of the free Hamiltonian
extends this domain such that the state $\Psi_0$ is included.
Although the wave function $\Psi_0$ is not in the domain of $H_0$,
the eigenfunction $\psi_0$ corresponding to the lowest eigenvalue
of $\Phi(E)$ can be taken in the domain of $H_0$ (see the
discussions in \cite{caglarermanturgut}).

As a result,  $|\Psi_0
\rangle $ given in (\ref{psi1}) is not in the product form in the
large $n$ limit, that is,
\be |\Psi_0 \rangle \neq \int_{\mathcal{M}^{n}} \mathrm{d}_{g}^{2}
y_1 \cdots \mathrm{d}_{g}^{2} y_n \; \prod_{k=1}^{n}
\Psi_{0}(y_{k}) |y_{1}\cdots y_{n} \rangle \;. \ee
The solution takes a kind of  convolution of the wave functions in
the domain of $H_0$ with the bound state wave function which is
outside of this domain.

Yet, $\Phi(E)$'s lowest eigenfunction may be approximated by a
product form for large number of bosons, that is,
\be \psi_{0}(x_1,\cdots,x_{n-1})= u(x_1)\cdots u(x_{n-1}) \ee
with the normalization
\begin{equation} \label{normalization}
||u ||^2 = \int_{\mathcal{M}} \mathrm{d}_{g}^{2} x \; |u(x)|^2 = 1
\;.
\end{equation}
In our two dimensional problem, for non-compact case,  $\chi = O(\ln n)$ as we
will see, so the product formula (\ref{psi1}) may not be a good
approximation. Similarly, in three dimensions for the noncompact case, $\chi=O(1)$.
In such a case, $u(x)$'s are not the wave function
of bosons when a bound state forms, but related to the correct
wave function through (\ref{psi0}). In fact, the full wave
function of the ground state could be read from
\be \mathbb{P}_0 = \left(%
\begin{array}{c}
  | \Psi_0 ^{(n+1)}\rangle \\
  |\omega_0^{(n)} \rangle \\
\end{array}%
\right) \otimes \left(%
\begin{array}{c}
  | \Psi_0^{(n+1)} \rangle \\
  |\omega_0^{(n)} \rangle \\
\end{array}%
\right)^{T} \;, \ee
where we explicitly denote the boson numbers. We call the
eigenvector of the principal operator in the mean field
approximation $|u \rangle$ for consistency of notation with our
previous paper \cite{ermanturgut2}. In the mean field
approximation, the operators are usually approximately replaced by
their expectation values in this state i.e., $\langle f(x) \rangle
\approx f(\langle x \rangle)$. However, the exact value of the
expectation value of an operator is given in terms of cummulant
expansion theorem if it converges \cite{Ma}. Therefore, we assume
that the corrections coming from the higher order cummulants are
negligibly small and indeed we will see that this assumption is
justified for the particular solution we will find. Therefore, the
expectation value of the principal operator by applying the mean
field ansatz becomes
\begin{eqnarray}
\phi_E[u]&=&n h_0[u]-E+\mu+\lambda^2\int_0^\infty \mathrm{d} s \;
K_{s}(a,a;g)\,[e^{-s(m-\mu)}-e^{-s(nh_0[u]+m-E)}]\cr &\ &
\hskip-2.5cm - \; n \lambda^2 \int_0^\infty \mathrm{d} s
\int_{\mathcal{M}^2} \mathrm{d}_{g}^{2} x \, \mathrm{d}_{g}^{2} y
\; K_{s}(x,a;g)\, K_{s}(y,a;g)\, u^*(x)\, e^{-s(nh_0[u]+2m-E)} \,
u(y) \;, \label{principalfunc}
\end{eqnarray}
and
\begin{equation}
h_0 [u]= \int_{\mathcal{M}} \mathrm{d}_{g}^{2} x \; \left( {
|\nabla_g u(x)|^2 \over 2 m} + m |u(x)|^2 \right) = K[u] + m \;,
\end{equation}
where we have taken $(n-1)\approx n$ for $n\gg 1$ and $K[u]$ is
called the kinetic energy functional. Now, we must solve the
functional equation $\phi_E[u] =0$ (giving the bound state
spectrum of the problem), that is, we solve $E$ as a functional of
$u(x)$, and then find the smallest possible value of $E$ with the
constraint (\ref{normalization}). One can try to write $E$ as a
functional of $u(x)$ from the equation $\phi_E[u] =0$ and apply
the variational methods to minimize $E=E[u]$. However, this is a
implicit function of $E$ and there is no simple way to solve
exactly this functional equation since $E$ is a complicated
functional of $u(x)$. Moreover,  we have no explicit expression of
the heat kernel on any Riemannian manifold to solve $E$.
Nevertheless, it is possible to find a lower bound on the ground
state energy without applying the variational calculus techniques.

Since $n h_0[u]$ and $E$ come together in equation
(\ref{principalfunc}), it turns out to be convenient to introduce
a new variable $\chi=\chi[u]$
\begin{eqnarray} \chi & \equiv & n h_0[u]-E \;.
\label{chi} \end{eqnarray}
Then, the condition $\phi_E[u]=0$ gives
\beqs & & \hskip-1.5cm \chi+\mu+\lambda^2\int_0^\infty \mathrm{d}
s \; K_{s}(a,a;g)\,[e^{-s(m-\mu)}-e^{-s(\chi+m)}]\cr &\ &
\hskip-0.5cm = n \lambda^2 \int_0^\infty \mathrm{d} s
\left|\int_{\mathcal{M}} \mathrm{d}_{g}^{2} x \; K_{s}(x,a;g)\,
u(x)\right|^2 \; e^{-s(\chi+2m)} \;. \label{principalfunc2} \eeqs
Note that the left hand side is an increasing function of $\chi$
while the right hand side is a decreasing function of $\chi$,
hence there is a unique solution for $\chi$.

To get a feel for the problem, we consider $\chi$ as the dependent variable.
We now remove the  $\chi$ dependence of the  right  hand
side of (\ref{principalfunc2}) by first defining a new
dimensionless parameter $s' = 2m(2m+\chi)s$ and scaling the metric
$\tilde g_{ij}=[2m(2m+\chi)] g_{ij}$. Using the scaling property
of heat kernel in two dimensions
\be K_{s}(x,y;g)=\alpha^{2} K_{\alpha^2 s}(x,y;\alpha^{2} g)
\;,\label{heatkernelscaling2} \ee
and then defining new dimensionless wave function $v(x)$
\begin{eqnarray}
v(x)&\equiv&[2m(2m+\chi)]^{-1/2}u(x) \;,
\end{eqnarray}
all explicit $\chi$ dependence becomes shifted to the left hand
side of (\ref{principalfunc2}). The condition $\phi_E[u]=0$ in two
dimensions then gives
\begin{eqnarray} & &
\Big(\chi+\mu+\lambda^2\int_0^\infty \mathrm{d} s \;
K_{s}(a,a;g)\left[e^{-s(m-\mu)}-e^{-s(\chi+m)}\right]\Big) \cr & \
&  = n\lambda^2(2m) \int_0^\infty \mathrm{d} s' \; \left|
\int_{\mathcal{M}} \mathrm{d}_{\tilde g}^{2} x \;
K_{s'}(x,a;\tilde g) \, v(x)\right|^2 e^{-s'/2m} \;.
\label{meanfieldeq2D}
\end{eqnarray}
Of course, $K_{s'}(x,a;\tilde g)$ has now a dependence on $\chi$
but let us assume that by varying $v(x)$ we can get all possible
values of the argument. It is important to notice that the left
hand side is an increasing function of $\chi$ and the right hand
side is always positive. Therefore, the left hand side is minimum
when $\chi=-\mu$. Let us denote the inverse function of the left
hand side as $f_1(nU)$, that is, $\chi= f_1(nU[v])$. Here $U[v]$
denotes the functional on the right hand side except for the
factor $n$. We can express $E$ in terms of the inverse function,
\be E= n m + 2mn K[v] +(n K[v]-1) f_1(n U)\;, \ee
where $K[v]=\int_{\mathcal{M}}\mathrm{d}^{2}_{g} x \; |\nabla_g
v(x)|^2$ is the dimensionless kinetic energy functional. Hence
unless $nK[v]<1$, the energy is always bigger than $nm+\mu$. As a
result, we see that the interesting possibility corresponds to
$nK[v]<1$ case. In the analysis for $nK[v]<1$, if we follow the
same reasoning as in three dimensions, the following integral
appears
\be \int_{0}^{\infty} \mathrm{d}s  \;
(1-e^{-s(\chi+2m)/2m})K_s(a,a;g) \;. \ee
However, an upper bound to this integral can not be found by using
the diagonal upper bounds of the heat kernel given in the previous
section because it is divergent for large values of $s$.
Therefore, we must develop a different method to handle the two
dimensional problem.

For the case $nK[v]<1$, we will again consider the problem for
compact and noncompact manifolds separately. Using the
eigenfunction expansion for the heat kernel \cite{ermanturgut2}
and for $v(x)= \sum_{l=0}^{\infty} v(l)f_l(x;g)$ and taking the
integral with respect to $s'$, we find the right hand side of
(\ref{meanfieldeq2D})
\beqs n (2m) \lambda^2 \sum_{l_1 = 0}^{\infty} \sum_{l_2 =
0}^{\infty} {1 \over 1+ \bar{\sigma}_{l_1} + \bar{\sigma}_{l_2}}
v^{*}(l_1) v(l_2) f_{l_1}(a;\tilde{g}) f_{l_2 }^{*}(a;\tilde{g})
\;, \label{rhseigenfunc} \eeqs
where $f_l(x;\tilde{g})$ and $\bar{\sigma}_l$ are the
eigenfunctions and the eigenvalues of the scaled Laplacian $-
\nabla_{\tilde{g}}^{2}$, respectively. Since this is always
positive, it is smaller than the following terms by writing the
zero modes separately
\beqs & & \leq n (2m) \lambda^2 \Bigg( |f_0(a;\tilde{g})|^2
|v(0)|^2 + 2 \Bigg|\sum_{\substack{l_1 \neq 0 \\ (l_2=0)}}
{f_{0}^{*}(a;\tilde{g}) v^{*}(0) f_{l_1}(a;\tilde{g}) v(l_1) \over
(1 + \bar{\sigma}_{l_1})} \Bigg| \cr & & \hspace{7cm} + \Bigg|
\sum_{l_1 \neq 0} \sum_{l_2 \neq 0} {f_{l_1}^{*}(a;\tilde{g})
v^{*}(l_1) f_{l_2}(a;\tilde{g}) v(l_2) \over (1 +
\bar{\sigma}_{l_1} + \bar{\sigma}_{l_2} )} \Bigg| \Bigg) \;. \eeqs
Using $f_0(a;\tilde{g}) = {1 \over
\sqrt{V(\mathcal{M}(\tilde{g}))}} $ and $|v(0)|\leq 1$ we get
\beqs & & \leq n (2m) \lambda^2 \Bigg( {1 \over
V(\mathcal{M}(\tilde{g}))}+ {2 \over \sqrt{V(\mathcal{M}(\tilde{g}))} }
\Bigg|\sum_{\substack{l_1 \neq 0 \\
(l_2=0)}} {f_{l_1}(a;\tilde{g}) v(l_1) \over (1 +
\bar{\sigma}_{l_1})} \Bigg| \cr & & \hspace{7cm} + \Bigg|\sum_{l_1
\neq 0} \sum_{l_2 \neq 0}{f_{l_1}^{*}(a;\tilde{g}) v^{*}(l_1)
f_{l_2}(a;\tilde{g}) v(l_2) \over (1 + \bar{\sigma}_{l_1} +
\bar{\sigma}_{l_2} )} \Bigg| \Bigg) \;. \label{meanfieldcompact2D}
\eeqs
Let us first consider the second term and multiply both numerator
and denominator with the factor $\bar{\sigma}_{l_1}^{1-\epsilon
\over 2}$ and then apply Cauchy-Schwartz inequality so that we
find
\beqs & & \sum_{l_1 \neq 0} {f_{l_1}(a;\tilde{g}) v(l_1) \over (1
+ \bar{\sigma}_{l_1})} \leq \Bigg(\sum_{l_1 \neq 0} |v(l_1)|^2
\bar{\sigma}_{l_1}^{1-\epsilon} \Bigg)^{1/2} \Bigg(\sum_{l_1 \neq
0} {|f_{l_1}(a;\tilde{g})|^2 \over (1 + \bar{\sigma}_{l_1})^2
\bar{\sigma}_{l_1}^{1-\epsilon} }\Bigg)^{1/2} \;, \label{2nd term}
\eeqs
where we have chosen $0< \epsilon < 1/2$. In order to convert the
products $(1 + \bar{\sigma}_{l_1})^2
\bar{\sigma}_{l_1}^{1-\epsilon}$ in the denominator into a
summation of them, we use a Feynman parametrization
\beqs {1 \over (1 + \bar{\sigma}_{l_1})^2
\bar{\sigma}_{l_1}^{1-\epsilon} }  &=& {\Gamma(3-\epsilon) \over
\Gamma(2)\Gamma(1-\epsilon)} \int_{0}^{1} \mathrm{d} u_1 \; u_1
\int_{0}^{1} \mathrm{d} u_2
 \; {\delta(u_1+u_2-1) u_{2}^{-\epsilon} \over (u_1(1 + \bar{\sigma}_{l_1}) + u_2
 \bar{\sigma}_{l_1})^{3-\epsilon}} \cr & = & {\Gamma(3-\epsilon) \over
\Gamma(2)\Gamma(1-\epsilon)} \int_{0}^{1} \mathrm{d} u_1 \;
{u_1(1-u_1)^{-\epsilon}  \over (u_1 +
\bar{\sigma}_{l_1})^{3-\epsilon} } \;. \label{feynman
parametrization1} \eeqs
One can express the factor ${1  \over (u_1 +
\bar{\sigma}_{l_1})^{3-\epsilon}}$ as an integral of
$s^{2-\epsilon}e^{-s (u_1 + \bar{\sigma}_{l_1})}$ due to
\be {1 \over a^{k+1}} = {1 \over \Gamma(k+1)} \int_{0}^{\infty}
\mathrm{d} s \; s ^k \; e^{-a s} \;, \label{expintegral} \ee
where $\Re(a)>0$ and $\Re(k)>-1$. Therefore,  equation
(\ref{feynman parametrization1}) becomes
\beqs {1 \over (1 + \bar{\sigma}_{l_1})^2
\bar{\sigma}_{l_1}^{1-\epsilon} }  & = & {1 \over
\Gamma(1-\epsilon)} \int_{0}^{1} \mathrm{d} u_1 \;
u_1(1-u_1)^{-\epsilon} \int_{0}^{\infty} {\mathrm{d} s' \over
2m}\; (s'/2m) ^{2-\epsilon} \; e^{-s'(u_1+\bar{\sigma}_{l_1})/2m}
\;. \eeqs
Using the eigenfunction expansion of the heat kernel, we have
\beqs & & \sum_{l_1 \neq 0} {|f_{l_1}(a;\tilde{g})|^2 \over (1 +
\bar{\sigma}_{l_1})^2 \bar{\sigma}_{l_1}^{1-\epsilon} } = {1 \over
\Gamma(1-\epsilon)} \int_{0}^{1} \mathrm{d} u_1 {1 \over
(1-u_1)^{\epsilon}} \int_{0}^{\infty} {\mathrm{d} s' \over 2m}\;
(s'/2m) ^{2-\epsilon} \; e^{-s'u_1/2m} \cr & & \hspace{9cm} \times
\Bigg(K_{s'}(a,a;\tilde{g})-{1 \over
V(\mathcal{M}(\tilde{g}))}\Bigg) \;. \eeqs
After we transform the above integral to the original metric $g$
and the time $s$ and take the $s$ integral by using the diagonal
upper bound of the heat kernel for compact manifolds (\ref{compact
upper bound heat kernel}), we find a upper bound of the above sum
\be \sum_{l_1 \neq 0} {|f_{l_1}(a;\tilde{g})|^2 \over (1 +
\bar{\sigma}_{l_1})^2 \bar{\sigma}_{l_1}^{1-\epsilon} } \leq {A
\Gamma(2-\epsilon) \over \Gamma(1-\epsilon)} \int_{0}^{1}
\mathrm{d} u_1 \; {u_{1}^{\epsilon-1} \over (1-u_1)^{\epsilon}} =
{A \pi \Gamma(2-\epsilon) \over \Gamma(1-\epsilon) \sin \pi
\epsilon}  \;. \ee
We now come to the crucial point. The upper bound of the first sum
in (\ref{2nd term}) can be found by using the following inequality
(also used in \cite{ali}),
\be \bar{\sigma}_{1}^{1-\epsilon} < \delta + \left({\epsilon \over
\delta}\right)^{\epsilon/1-\epsilon} \bar{\sigma}_{1}
\label{inequality} \;,\ee
where $\delta> 0 $ and $0< \epsilon < 1/2$. Note that this
inequality applies only to dimensionless variables
$\bar{\sigma}_{1}$, so that is why we use the scaling
transformation at the beginning of the problem as opposed to the
three dimensional case. The proof of this inequality is given in
Appendix. As a result, for the first sum of (\ref{2nd term}), we
obtain
\be \sum_{l_1 \neq 0} |v(l_1)|^2 \bar{\sigma}_{l_1}^{1-\epsilon}
 \leq \delta + \left({\epsilon \over
\delta}\right)^{\epsilon/1-\epsilon} K[v]\;, \label{thirdterm
compact mf} \ee
where excluding the zero mode from the sum again gives the kinetic
energy functional $K[v]$ since $\bar{\sigma}_{0}=0$. Then, we
finally get
\beqs  & & {2 n (2m) \lambda^2 \over \sqrt{V(\mathcal{M}(\tilde{g}))} }
\Bigg|\sum_{\substack{l_1 \neq 0 \\
(l_2=0)}} {f_{l_1}(a;\tilde{g}) v(l_1) \over (1 +
\bar{\sigma}_{l_1})} \Bigg| \cr & & \hspace{2cm} \leq {2n (2m)
\lambda^2 \over \sqrt{2m(2m+\chi)V(\mathcal{M}(g))} }\Bigg[\delta
+ \left({\epsilon \over \delta}\right)^{\epsilon/1-\epsilon}
K[v]\Bigg]^{1/2} \Bigg[{A \pi \Gamma(2-\epsilon) \over
\Gamma(1-\epsilon) \sin \pi \epsilon} \Bigg]^{1/2}
\label{secondterm mf2} \;. \eeqs
Since $\sin \pi \epsilon \geq 2 \epsilon $ for $0\leq \pi\epsilon
\leq \pi/2$ (a useful inequality: $\sin \theta \geq 2 \theta/\pi$
for $0 \leq \theta \leq \pi/2$) and $nK[v]<1$, the last expression
is smaller than
\beqs  {2 \sqrt{n} \sqrt{2m} \lambda^2 \over
\sqrt{(2m+\chi)V(\mathcal{M}(g))} }\Bigg[n \left( {\delta \over
\epsilon}\right) + {1 \over \epsilon} \left({\epsilon \over
\delta}\right)^{\epsilon/1-\epsilon} \Bigg]^{1/2} \Bigg[{A \pi
\Gamma(2-\epsilon) \over 2 \Gamma(1-\epsilon)} \Bigg]^{1/2}
\label{secondterm 3} \;. \eeqs
By choosing arbitrary constants $\epsilon$ and $\delta$, we see
that the interaction term brings a contribution of order $O(n)$ to
the total energy. However, one can even find a better solution to
the large $n$ behavior of the energy. In order to control the
energy as $n\rightarrow \infty$, we can assume that the parameters
$\epsilon$ and $\delta$ are sequences in $n$. Without loss of
generality we can assume that $\epsilon(n)$ goes to zero as
$n\rightarrow \infty$ (recall that $0<\epsilon<1/2$). If we want
to find a better large $n$ behavior of the energy, we must choose
the sequence $\delta(n)$ such that
\be {n \delta(n) \over \epsilon^2(n)} = O(1)\;. \label{epsilondelta
order} \ee
We are now looking for an optimal solution for the energy and tell
how fast the sequences $\epsilon(n)$ and $\delta(n)$ must change
with $n$. In order to see this, let us write (\ref{secondterm 3})
in the following way
\beqs
 {2 \sqrt{n} \sqrt{2m} \lambda^2 \over
\sqrt{(2m+\chi)V(\mathcal{M}(g))} }  \Bigg[{n \delta(n) \over
\epsilon(n)} + e^{{\epsilon(n) \over 1-\epsilon(n)}\ln
(\epsilon(n)/\delta(n))-\ln \epsilon(n)} \Bigg]^{1/2} \Bigg[{A \pi
\Gamma(2-\epsilon(n)) \over 2 \Gamma(1-\epsilon(n))} \Bigg]^{1/2}
\;. \label{thirdterm 4} \eeqs
An optimal solution of the sequences can be found in such a way
that the exponential term goes asymptotically
\beqs e^{{\epsilon(n) \over 1-\epsilon(n)}\ln
(\epsilon(n)/\delta(n))-\ln \epsilon(n)} =O(\ln n) \;. \eeqs
This implies that we can choose
\beqs \epsilon(n) = {1 \over \ln n}\;, \eeqs
and as a consequence of (\ref{epsilondelta order})
\beqs \delta(n) = {1 \over n \ln^2 n} \;.\eeqs
Therefore, we obtain the upper bound of (\ref{secondterm mf2}) for
$n \gg 1$
\beqs & & {2 n (2m) \lambda^2 \over \sqrt{V(\mathcal{M}(\tilde{g}))} }
\Bigg|\sum_{\substack{l_1 \neq 0 \\
(l_2=0)}} {f_{l_1}(a;\tilde{g}) v(l_1) \over (1 +
\bar{\sigma}_{l_1})} \Bigg| \leq {\lambda^2 \sqrt{4 m A \pi e}
\over {\sqrt{(2m+\chi) V(\mathcal{M})}}} \sqrt{n \ln n} \;.
\label{secondterm compact mf result}\eeqs
As for the last term in (\ref{meanfieldcompact2D}), the procedure
outlined above is very similar. However, we now multiply both
numerator and denominator of this term with the factor
$\bar{\sigma}_{l_1}^{1-\epsilon \over 2}
\bar{\sigma}_{l_2}^{1-\epsilon \over 2}$ and apply Cauchy-Schwartz
inequality and get
\beqs & & \sum_{l_1 \neq 0} \sum_{l_2 \neq
0}{f_{l_1}^{*}(a;\tilde{g}) v^{*}(l_1) f_{l_2}(a;\tilde{g}) v(l_2)
\over (1 + \bar{\sigma}_{l_1} + \bar{\sigma}_{l_2} )} \cr & &
\hspace{3cm} \leq \Bigg[\sum_{l_1 \neq 0} |v(l_1)|^2
\bar{\sigma}_{l_1}^{1-\epsilon} \Bigg]  \Bigg[\sum_{l_1 \neq 0}
\sum_{l_2 \neq 0} {|f_{l_1}(a;\tilde{g})|^2
|f_{l_2}(a;\tilde{g})|^2 \over (1 + \bar{\sigma}_{l_1} +
\bar{\sigma}_{l_2} )^2 \bar{\sigma}_{l_1}^{1-\epsilon}
\bar{\sigma}_{l_2}^{1-\epsilon} } \Bigg]^{1/2} \;. \label{compact
2d mean field upper bound eq} \eeqs
For simplicity, we can use the following inequality in the second
sum ${1 \over (1 + \bar{\sigma}_{l_1} + \bar{\sigma}_{l_2})^2}
\leq {1 \over (1 + \bar{\sigma}_{l_1})(1 + \bar{\sigma}_{l_2})}$
since $\bar{\sigma}_{l_0}=0$ and obtain an upper bound on
(\ref{compact 2d mean field upper bound eq})
\beqs \leq \Bigg[\sum_{l_1 \neq 0} |v(l_1)|^2
\bar{\sigma}_{l_1}^{1-\epsilon} \Bigg]  \Bigg[ \sum_{l_1 \neq 0}
{|f_{l_1}(a;\tilde{g})|^2 \over (1 + \bar{\sigma}_{l_1})
\bar{\sigma}_{l_1}^{1-\epsilon}}\Bigg] \label{prethirdterm
compactmf} \;.  \eeqs
We again convert the products $(1 + \bar{\sigma}_{l_1})
\bar{\sigma}_{l_1}^{1-\epsilon}$ in the denominator into a
summation of them by using a Feynman parametrization
\be {1 \over (1 + \bar{\sigma}_{l_1})
\bar{\sigma}_{l_1}^{1-\epsilon}} = {\Gamma(2-\epsilon) \over
\Gamma(1-\epsilon)} \int_{0}^{1} \mathrm{d} u_1 {1 \over
(1-u_1)^{\epsilon} (u_1 + \bar{\sigma}_{l_1})^{2-\epsilon}} \;.
\ee
Then, we rewrite the factor ${1 \over (u_1 +
\bar{\sigma}_{l_1})^{2-\epsilon}}$ using (\ref{expintegral}) and
get
\beqs \sum_{l_1 \neq 0} {|f_{l_1}(a;\tilde{g})|^2 \over (1 +
\bar{\sigma}_{l_1}) \bar{\sigma}_{l_1}^{1-\epsilon}} &=& {1 \over
\Gamma(1-\epsilon)} \int_{0}^{1} \mathrm{d} u_1 {1 \over
(1-u_1)^{\epsilon}} \int_{0}^{\infty} {\mathrm{d} s' \over 2m}
(s'/2m)^{1-\epsilon} \cr & & \hspace{2cm} \times \sum_{l_1 \neq 0}
e^{-s'(u_1+\bar{\sigma}_{l_1})/2m}|f_{l_1}(a;\tilde{g})|^2 \;.
\label{compact2deq} \eeqs
Using the eigenfunction expansion of the heat kernel above leads
to
\beqs \hskip-0.5cm {1 \over \Gamma(1-\epsilon)} \int_{0}^{1}
\mathrm{d} u_1 {1 \over (1-u_1)^{\epsilon}} \int_{0}^{\infty}
{\mathrm{d} s' \over 2m} (s'/2m)^{1-\epsilon} e^{-s' u_1/2m}
\Bigg(K_{s'}(a,a;\tilde{g})-{1 \over
V(\mathcal{M}(\tilde{g}))}\Bigg) \;. \eeqs
Scaling back to the original variables and using the diagonal
upper bound of the heat kernel for compact manifolds (\ref{compact
upper bound heat kernel}), and taking the $s$ integral, we obtain
an upper bound of (\ref{compact2deq})
\beqs  A \int_{0}^{1} \mathrm{d} u_1 {u_1^{\epsilon-1} \over
(1-u_1)^{\epsilon}} = {\pi A \over \sin \pi \epsilon} \;.
 \eeqs
The analysis for the first sum in (\ref{prethirdterm compactmf})
is exactly the same as before. Therefore, the upper bound of the
third term in (\ref{meanfieldcompact2D}) for $n \gg 1$
\beqs & & \hskip-2cm n (2m)  \lambda^2 \Bigg|\sum_{\substack{l_1 \neq 0 \\
l_2 \neq 0}} {f_{l_1}^{*}(a;\tilde{g}) v^{*}(l_1)
f_{l_2}(a;\tilde{g}) v(l_2) \over (1 + \bar{\sigma}_{l_1} +
\bar{\sigma}_{l_2} )} \Bigg|  \leq 2m A \lambda^2 \pi e \ln n \;.
\label{thirdterm compact mf result}\eeqs
Combining the upper bounds (\ref{secondterm compact mf result})
and (\ref{thirdterm compact mf result}), we finally obtain an
upper of (\ref{meanfieldcompact2D}) for $n \gg 1$
\beqs \Bigg[{\lambda^2 \over (2m+ \chi)V(\mathcal{M}(g))}\Bigg]n +
\lambda^2 \left[4 m \pi e A \over
(2m+\chi)V(\mathcal{M}(g))\right]^{1/2}  n^{1/2} \ln^{1/2}n +
\Bigg[2mA \lambda^2 \pi e \Bigg] \ln n \;.\eeqs
For the left hand side of (\ref{meanfieldeq2D}), we have
\be \chi+\mu \leq \chi+\mu+\lambda^2\int_0^\infty \mathrm{d} s \;
K_{s}(a,a;g)\left[e^{-s(m-\mu)}-e^{-s(\chi+m)}\right] \;,
 \ee
due to the positivity of the heat kernel and $\chi\geq -\mu$.
Using this result and the upper bound of the right hand side of
(\ref{meanfieldeq2D}), we have
\be \chi+\mu \leq \Bigg[{\lambda^2 \over (2m+
\chi)V(\mathcal{M}(g))}\Bigg]n + \lambda^2 \left[4 m \pi e A \over
(2m+\chi)V(\mathcal{M}(g))\right]^{1/2} n^{1/2} \ln^{1/2}n +
\Bigg[2mA \lambda^2 \pi e \Bigg] \ln n \;. \ee
We will solve $\chi$ from this inequality for the large values of
$n$. It is important to notice that the right hand side is a
monotonically decreasing and the left hand side is a monotonically
increasing function of $\chi$. Therefore, if we find a solution to
the above equation, say at $\chi=\chi_*$, we conclude that $\chi
\leq \chi_*$. It is sufficient to find the leading order term of
$\chi$ for our purposes so $\chi_*$ can be taken as an infinite
power series
\be \chi_* \sim a_1 n^{\alpha_1} + a_2 n^{\alpha_2} + \ldots \ee
with the decreasing power $\alpha_1> \alpha_2>\ldots$ of $n$ and
$a_1,a_2,\ldots$ are the coefficients which can be found by
substitution. Then we find the leading order term of $\chi$
\be \chi \leq {\lambda \over \sqrt{V(\mathcal{M})}} n^{1/2} \ee
for $n \gg 1$. As a result of this, we get the lower bound of the
bound state energy $E_{gr}$ for large values of $n$
\beqs & & E_{gr} \sim nm + \mu - {\lambda \over
\sqrt{V(\mathcal{M})}} n^{1/2} \;. \eeqs
As for the noncompact manifolds, the analog expression of
(\ref{rhseigenfunc}) is
\beqs n (2m) \lambda^2 \int \int \mathrm{d} \mu (l_1) \mathrm{d}
\mu(l_2)  {1 \over 1+ \bar{\sigma}_{l_1} + \bar{\sigma}_{l_2}}
v^{*}(l_1) v(l_2) f_{l_1}(a;\tilde{g}) f_{l_2}^{*}(a;\tilde{g})
\;,\label{rhseigenfunc2} \eeqs
where $d\mu$ here refers to the formal spectral measure. One can think that
the sums are replaced with the integrals in the case for
noncompact cases and the analysis is basically same as the one for
compact manifolds except that we do not have to bother for
extracting the zero mode. Following the same steps for the analog
expression of the compact cases, and using the diagonal upper
bound of the heat kernel on Cartan-Hadamard manifolds
(\ref{cartanestim}), the ground state energy goes like
\be E_{gr} \sim n m + \mu - 2m C e \lambda^2 \ln n \;. \ee
Incidentally, these show that $K[u]\sim \chi K[v]=y[v] \chi n^{-1}
\approx y[v] n^{-(1-\alpha)}$ where $\alpha=1/2$ for compact and
$\alpha=0$ for noncompact manifolds. Hence, the cummulant
approximation is expected to give only the term that we kept, and
this shows that the approximation is consistent.

The solution actually implies an ansatz for the solution which is
possible for compact manifolds. Let us set $u={1\over
\sqrt{V({\cal M})}}$. Then we find for the energy
\begin{eqnarray}
nm-E + \mu +\lambda^2\int_0^\infty \mathrm{d} s \;
K_{s}(a,a;g)\left[e^{-s(m-\mu)}-e^{-s(nm-E)}\right] =
n{\lambda^2\over V({\cal M})}\left({1\over nm-E}\right) \;.
\end{eqnarray}
Now using a scaling for $s\mapsto s'/(nm-E)$ and using the
diagonal asymptotic expansion (\ref{asymheat}) for $K_s(a,a;g)$,
\begin{eqnarray}
nm-E+\mu +\lambda^2\int_0^\infty \mathrm{d} s' \; {2m\over 4\pi
s'}\left[e^{-s'(m-\mu)/(nm-E)}-e^{-s'}\right]=n{\lambda^2\over
V({\cal M})} \left({1\over nm-E} \right) \;,
\end{eqnarray}
we find
\begin{eqnarray}
nm-E+\mu +{m \lambda^2\over 2\pi} \ln \left({nm-E\over
m-\mu}\right)
 =n{\lambda^2\over V({\cal M})} \left({1\over nm-E}\right) \;.
\end{eqnarray}
This has the asymptotic solution as claimed in the compact case.
Interestingly, the similar ansatz for the three dimensional
compact manifolds gives exactly the same type of asymptotics for
the energy as in two dimensions, which is better than the upper
bound obtained of  the mean field analysis (note that one also has
$K[u]\sim n^{-1/3}$ so it suggests a constant wave function as an
ansatz).

\section{Conclusion}

In this paper, we considered the non-relativistic Lee model on
various class of two dimensional Riemannian manifolds. It is just
a continuation of our previous work \cite{ermanturgut2}. The
construction of the model and the calculations for lower bound of
the ground state energy is almost the same as in three dimensions.
However, the mean field approximation in two dimensions is not as
simple as in three dimensions and it required more analysis to predict
more precisely the large particle  limit of the model for
compact and non-compact manifolds.

\section{Appendix: The Proof of the Inequality} \label{The Proof of the
Inequality}

In order to prove the following inequality
\be  x^{1-\epsilon} < \delta + \left({\epsilon \over
\delta}\right)^{{\epsilon \over 1-\epsilon}} x \;, \ee
where $x$ is a dimensionless variable, let us consider the
following polynomial function
\be f(x)= \delta + \left({\epsilon \over \delta}\right)^{{\epsilon
\over 1-\epsilon}} x - x^{1-\epsilon} \;. \ee
We assume that $\delta > 0$ and $0 < \epsilon < 1/2$. This
function has one extremum point at $x_{*}$
\be x_{*}^{\epsilon} =  (1-\epsilon) \left( {\delta \over
\epsilon} \right)^{{\epsilon \over 1-\epsilon}} \;, \ee
and this location $x_*$ corresponds to the minimum. One can easily
see that
\be f(x_*) = \delta \left(1-(1-\epsilon)^{{1-\epsilon \over
\epsilon}}\right) \;.  \ee
Since $\epsilon < 1/2$, we have ${1-\epsilon \over \epsilon} > 1$.
We can also show that $(1-\epsilon)^{\alpha} < 1-\epsilon $ for
$\alpha >1$ since $1-\epsilon <1$ or
\be (1-\epsilon)^{{1-\epsilon \over \epsilon}} < 1-\epsilon \;.
\ee
Then, we find
\be f(x_*) > \delta \epsilon  \;, \ee
and it is always positive. Since this is a global minimum point,
we obtain $f(x)>0$ for all $x$, which completes the proof.

\section{Acknowledgments}

The authors gratefully acknowledge the discussions of S. G.
Rajeev, \c{C}. Do\~{g}an, B. Kaynak . O. T. T. 's research is
partially supported by the Turkish Academy of Sciences, in the
framework of the Young Scientist Award Program
(OTT-TUBA-GEBIP/2002-1-18).

\end{document}